\documentclass{jpsj3}
\usepackage{graphicx}
\usepackage{dcolumn}
\usepackage{bm}
\usepackage{color}

\title{%
Two-Photon Absorption by Impurity Scattering and Amplitude Mode in Conventional Superconductors 
}
\author{%
Takanobu Jujo
\thanks{E-mail address: jujo@ms.aist-nara.ac.jp}
}
\inst{%
Graduate School of Materials Science, 
Nara Institute of Science and Technology, 
Ikoma, Nara 630-0101, Japan
}
\recdate{\today}
\abst{%
We theoretically investigate 
the two-photon absorption spectrum 
of conventional s-wave superconductors. 
The calculation is carried out 
in the local limit at absolute zero. 
The direct excitation of quasiparticles 
is excluded under the condition that 
the frequency of the external field is 
lower than twice the magnitude of 
the superconducting gap. 
We calculate vertex corrections 
with the conserving approximation, 
and show that the amplitude mode is 
independent of the impurity scattering in the local limit. 
Because of this property, 
a diverging absorption edge appears in the dirty limit. 
We propose this behavior as a probe of the amplitude mode. 
}

\begin{document}
\setlength{\textwidth}{504pt}
\setlength{\columnsep}{14pt}
\hoffset-23.5pt
\maketitle

\section{Introduction}

The superconducting transition is accompanied by 
collective modes.
~\cite{schon} 
Corresponding to degrees of freedom of 
the order parameter, 
there are phase and amplitude modes. 
The former couples with a density fluctuation 
and becomes a plasmon.~\cite{anderson1} 
This makes the experimental observation of the phase 
mode difficult. 
(The relative phase mode appears below the 
superconducting gap ($\Delta$) 
in two-band superconductors~\cite{leggett} 
or in the Josephson junction between two superconductors.~\cite{jujo12} 
This mode is used as a superconducting quantum bit.~\cite{clarke}) 

The amplitude mode~\cite{schmid,volkov} was investigated 
via phonon degrees of freedom 
in the superconducting charge-density-wave 
material~\cite{sooryakumar,littlewood} 
because this mode does not directly 
interact with the external field 
(the interaction term is not given in 
the bilinear form). 
Therefore, the objects of this investigation have been restricted 
within specific materials. 
The development of the terahertz technique made 
the amplitude mode observable in other materials. 
This technique induces nonlinear responses, and 
an investigation by 
pump-probe spectroscopy 
has recently been performed.~\cite{matsunaga}   

By this method, an oscillation of 
the conductivity was observed by exciting 
quasiparticles by the pump pulse with its frequency 
comparable to $2\Delta$. 
This type of observation of the transient state 
is a major method for studying 
the nonequilibrium superconductivity. 
In this paper, we show, however, that 
the amplitude mode is detectable 
in the two-photon absorption (TPA) spectrum 
without exciting extra degrees of freedom 
such as quasiparticles and phonons. 
Because a direct excitation of quasiparticles 
above the superconducting gap is excluded, 
this investigation can be performed in the steady state 
without destroying a superconducting state. 

The conditions for our calculation are as follows. 
Absorption spectra are calculated in the 
dirty and local limits, 
~\cite{note1}
which hold in a pump-probe experiment.~\cite{matsunaga} 
Except for a calculation of the linear response, 
the frequency of the external field ($\omega$) is 
set to be lower than $2\Delta$, 
and then there is no direct excitation of quasiparticles 
across the superconducting gap. 
We also consider the case of the low-temperature limit 
($T\to 0$) 
in which the thermal excitation of quasiparticles is 
excluded.
Under these conditions, 
the result of our calculation shows that 
the amplitude mode causes 
a divergent peak ($\sim 1/\sqrt{\omega-\Delta}$) 
around the absorption edge $\omega=\Delta$, 
which is quantitatively estimated to be 
detectable. 

In Sect. 2, the formulation to calculate the 
current density is given with a definition of 
the action. 
In Sect. 3, the one-particle Green function, 
the linear absorption function, and 
the TPA function without the vertex correction are calculated. 
In Sect. 4, the vertex correction is introduced 
and classified into the amplitude mode and 
impurity scattering. 
The TPA functions with these vertex corrections 
are calculated. 
In Sect. 5, numerical calculations are 
performed with use of the results in Sects. 3 and 4. 
In Sect. 6, a short summary is given. 
A brief discussion for a future study 
is also given. 
We set $\hbar=1$ in this paper.

\section{Formulation}

The response function is calculated with the use of 
the method of nonequilibrium Green's function.~\cite{keldysh,kamenev}  
The definition of the current density is written as 
\begin{equation}
 J_q(\omega)=
\frac{-i}{\sqrt{2}}
\frac{\delta \text{ln}Z[A]}{\delta A^{qu}_{-q}(-\omega)}
\Biggl|_{A^{qu}\to 0}. 
\label{eq:crnt}
\end{equation}
Here, $i=\sqrt{-1}$ and 
$Z[A]=\int{\cal D}[\psi,b]\text{e}^{iS_{\psi,b}[A]}$. 
The action $S_{\psi,b}[A]=S_{\psi}+S_b+S_{\psi\text{-}A}
+S_{\psi\text{-}b}+S_{imp}$ 
is given by the summation of terms for electrons ($\psi$) 
[$S_{\psi}=\int{\rm d}t\sum_{k,\sigma}
\bar{\psi}_{k,\sigma}(t)
(i\partial_t-\xi_k)\psi_{k,\sigma}(t)$ 
($k$ and $\sigma$ indicate the wave number and 
spin, respectively)], 
phonons ($b$) 
[$S_{b}=\int{\rm d}t\sum_{q}
\bar{b}_{q}(t)
(i\partial_t-\omega_q)b_{q}(t)$], 
the interaction between electrons and the external field ($A$)
[$S_{\psi\text{-}A}=\int{\rm d}t(N^3)^{-1/2}\sum_{k,q,\sigma}
\bar{\psi}_{k+q,\sigma}(t)
\text{e}A_q(t)v_{k+q/2}\psi_{k,\sigma}(t)$ with 
$A_q(t)$ as the vector potential], 
the electron-phonon interaction
[$S_{\psi\text{-}b}=\int{\rm d}tg_{ph}(N^3)^{-1/2}\sum_{k,q,\sigma}
\bar{\psi}_{k+q,\sigma}(t)
(b_q(t)+\bar{b}_{-q}(t))\psi_{k,\sigma}(t)$ 
($g_{ph}$ is the coupling constant)], 
and the impurity scattering for electrons 
[$S_{imp}=\int{\rm d}t(N^3)^{-1/2}\sum_{k,k',\sigma}
\bar{\psi}_{k,\sigma}(t)
u_{k-k'}\psi_{k',\sigma}(t)$]. 
($N^3$ is the number of sites. 
$\omega_q$ is the frequency of phonons. 
$\xi_k=k^2/2m-E_F$ is the kinetic energy 
of electrons measured from the Fermi energy $E_F$, 
and $v_k=\partial \xi_k/\partial k$. 
The direction of the external field is fixed 
and its index is omitted. 
$u_{k-k'}$ represents the impurity potential 
and will be assumed to be isotropic.) 

The integration ($\int{\rm d}t$) is 
carried out on the forward and backward paths 
in time.~\cite{kamenev} 
The Coulomb interaction is omitted because 
it is not effective except for pushing up 
the phase mode to the plasmon.~\cite{anderson1} 
$A^{qu}=(A^+ - A^-)/\sqrt{2}$, in which 
$A^{+}$ and $A^{-}$ represent 
the vector potentials in 
the forward and backward directions in time, respectively. 
The physical external field 
is given by $A^{cl}/\sqrt{2}$ 
with $A^{cl}=(A^+ + A^-)/\sqrt{2}$. 
(The definition of $A^{cl,qu}$ 
is different from that in Ref. 13 
%Ref.~\cite{kamenev} 
by a factor $\sqrt{2}$.) 

\section{Response Function}

\subsection{Linear response}

First, we show the calculation 
of the linear response, 
and in this case, the current density is written as 
$J^{(1)}_{\omega}=-K^{(1)}_{\omega}A_{\omega}$
with 
$K^{(1)}_{\omega}=\text{e}^2(4\pi i N^3)^{-1}
\sum_k\int\text{d}\epsilon
\text{Tr}
[\check{G}_{k}(\epsilon+\omega)
\check{G}_k(\epsilon)\check{1}^{qu}](v_{k})^2$. 
Here, 
$\check{1}^{qu}=\bigl(\begin{smallmatrix}
\hat{0} & \hat{1} \\ \hat{1} & \hat{0} \end{smallmatrix}\bigr)$ 
($\hat{1}$ and $\hat{0}$ are 
the unit and null $2\times 2$ matrix, respectively). 
In the calculation of the response function, 
we omit the diamagnetic term, 
which is proportional to 
$1/m=\partial v_k/\partial k$ (the inverse of the electron mass) 
because it is negligible.
~\cite{note2} 
(Hereafter, the index of the wave vector $q$ is omitted 
because we consider the local limit $q=0$.) 
The electronic Green's function in the Keldysh-Nambu space is 
\[
 \check{G}_k(\epsilon)=
\begin{pmatrix}
\hat{G}^+_k(\epsilon) & \hat{G}^{K}_k(\epsilon) \\ 
\hat{0} & \hat{G}^{-}_k(\epsilon) 
\end{pmatrix}. 
\]
This is calculated 
with the Born approximation for the impurity scattering~\cite{abrikosov} 
and the weak coupling approximation for the electron-phonon 
interaction, and the results are 
\[
 \hat{G}^{\pm}_k(\epsilon)=
\frac{\eta^{\pm}_{\epsilon}\epsilon\hat{1}+\xi_k\hat{\tau}_3
+\eta^{\pm}_{\epsilon}\Delta\hat{\tau}_1
}{(\eta^{\pm}_{\epsilon}\epsilon)^2-
\xi_k^2-(\eta^{\pm}_{\epsilon}\Delta)^2} 
\] 
and 
$\hat{G}^{K}_k(\epsilon)=
\text{tanh}(\epsilon/2T)
[\hat{G}^{+}_k(\epsilon)-\hat{G}^{-}_k(\epsilon)]$. 
Here, 
$\hat{\tau}_1=\bigl(\begin{smallmatrix}
0 & 1 \\ 1 & 0 \end{smallmatrix}\bigr)$, 
$\hat{\tau}_2=\bigl(\begin{smallmatrix}
0 & -i \\ i & 0 \end{smallmatrix}\bigr)$, 
and $\hat{\tau}_3=\bigl(\begin{smallmatrix}
1 & 0 \\ 0 & -1 \end{smallmatrix}\bigr)$ 
are 
Pauli matrices, and 
$\eta^{\pm}_{\epsilon}=
1+i\alpha/X_{\pm\epsilon}$ 
with 
$X_{\epsilon}=\text{sgn}(\epsilon)\sqrt{\epsilon^2-\Delta^2}
\theta(|\epsilon|-\Delta)
+i\sqrt{\Delta^2-\epsilon^2}\theta(\Delta-|\epsilon|)$ and 
$\alpha=n_iu^2mk_F/2\pi$ 
\{
$n_i$ and $u$ are the concentration of impurities and 
the magnitude of the impurity potential, respectively. 
$k_F$ is the Fermi wave number. 
$\text{sgn}(\epsilon)=\epsilon/|\epsilon|$, 
and $\theta(x)=\lim_{T\to 0}[1+\text{tanh}(x/2T)]/2$ 
is a step function\}. 
We consider the Einstein phonon with its frequency 
$\omega_q=\omega_E$, and 
in the weak coupling approximation ($\Delta\ll\omega_E$), 
~\cite{tsuneto} 
the gap equation 
is written as 
\[
 \Delta\hat{\tau}_1=
\frac{g_{ph}^2}{\omega_E}
\frac{1}{N^3}\sum_{k}\int\frac{\text{d}\epsilon}{2\pi i}
\hat{\tau}_3\hat{G}^{K}_{k}(\epsilon)\hat{\tau}_3.
\]

The calculated result for the linear absorption spectrum 
(the real part of the conductivity; 
$\text{Re}\sigma^{(1)}_{\omega}=-\text{Im}K^{(1)}_{\omega}/\omega$) 
is written as 
\begin{equation}
 \text{Re}\sigma^{(1)}_{\omega}=\sigma_{\Delta}
\frac{\alpha\Delta}{3\pi^2\omega}
\int_{\Delta-\omega/2}^{-\Delta+\omega/2}\text{d}\epsilon
\sum_{s=\pm 1}\frac{s+
(\epsilon^2-\omega^2/4+\Delta^2)
/(X_{\epsilon+\omega/2}X_{\epsilon-\omega/2})}
{(X_{\epsilon+\omega/2}-sX_{\epsilon-\omega/2})^2+4\alpha^2}
\theta(\omega-2\Delta), 
\label{eq:linab}
\end{equation}
which is the same as that of 
previous works.~\cite{mattis,zimmermann}
Here, $\sigma_{\Delta}=\text{e}^2k_F^3/m\Delta$, and 
the relaxation time $\tau$ in Ref. 18 
%Ref.~\cite{zimmermann} 
corresponds to $1/2\alpha$ in our calculation, 
and its value is roughly $\alpha\simeq 50\sim 70$ in Ref. 10. 
%Ref.~\cite{matsunaga}. 
In the local limit, 
the absorption occurs owing to 
the finite values of $\alpha$ and 
increases with $\alpha$ 
for small values of $\alpha$ ($\alpha\lesssim\Delta$). 
For large values of $\alpha$ ($\alpha\gg\Delta$), 
\[
\text{Re}\sigma^{(1)}_{\omega} \simeq
\sigma_0
\int_{\Delta-\omega/2}^{-\Delta+\omega/2}\text{d}\epsilon
\frac{\epsilon^2-\omega^2/4+\Delta^2}
{\omega X_{\epsilon+\omega/2}X_{\epsilon-\omega/2}}
\theta(\omega-2\Delta). 
\]
Here, $\sigma_0=\text{e}^2n_e\tau/m$ 
with $n_e=k_F^3/3\pi^2$ and $\tau=1/2\alpha$. 
Then, the absorption 
decreases with increasing $\alpha$ 
and $\text{Re}\sigma^{(1)}\propto 1/\alpha$. 
It is shown below that 
there is a difference in the absorption edge 
between the linear and nonlinear responses. 
The absorption goes to zero for $\omega\to2\Delta$ 
(a vanishing absorption edge), as known from 
the above expression for $\text{Re}\sigma^{(1)}_{\omega}$. 
This originates from a property of the vertex 
$v_k\hat{1}$. 
The vertex correction 
is not effective in the linear response 
because of the inversion symmetry, 
as discussed in Sect. 4.1.

\subsection{Third-order response without the vertex correction}

The third-order nonlinear response function 
is introduced 
to calculate the TPA spectrum. 
The current density in this case is given by 
$J^{(3)}_{\omega}=-K^{(3)}_{\omega}A_{\omega}$ 
with 
$K^{(3)}_{\omega}=K^{(3nv)}_{\omega}+K^{(3vc)}_{\omega}$. 
($K^{(3nv)}_{\omega}$ and $K^{(3vc)}_{\omega}$ are defined below.) 
The third-order response function with no vertex correction 
is $K^{(3nv)}_{\omega}=
2(
\tilde{K}_{\omega,\omega,-\omega}+
\tilde{K}_{\omega,-\omega,\omega}+
\tilde{K}_{-\omega,\omega,\omega})
|E|^2/\omega^2$ 
with $\tilde{K}_{\omega_1,\omega_2,\omega_3}
=
\text{e}^4(4\pi i N^3)^{-1}
\sum_k\int\text{d}\epsilon
\text{Tr}[
\check{G}_{k}(\epsilon+\omega_1+\omega_2+\omega_3)
\check{G}_{k}(\epsilon+\omega_2+\omega_3)
\check{G}_{k}(\epsilon+\omega_3)
\check{G}_{k}(\epsilon)\check{1}^{qu}] (v_k)^4$. 
($|E|$ is the magnitude of the electric field.) 
The absorption spectrum obtained 
by this process is given by 
\begin{equation}
\text{Re}\sigma^{(3nv)}_{\omega}=
\sigma_{\Delta}
\left(\frac{\text{e}\xi_0|E|}{\omega}\right)^2
\frac{\Delta^3}{5\pi^2\omega}
\int_{\Delta-\omega}^{-\Delta+\omega}\text{d}\epsilon
\sum_{s=\pm 1}s\text{Re}
\left(
\frac{F^s_{A}+F^s_{AB}+F^s_{B}+2F^s_{C}}
{z_0(z_0+z_1)^2(z_0+z_2)^2(z_1+z_2)}
\right)\theta^<_{\omega}. 
\label{eq:novcab}
\end{equation}
Here, 
$\theta^<_{\omega}=\theta(\omega-\Delta)\theta(2\Delta-\omega)$ 
(the first and second $\theta$'s correspond to 
the absence of 
thermal excitation and a restriction for our calculation, 
respectively), 
$\xi_0=v_F/\Delta$ ($v_F=k_F/m$ is the Fermi velocity), 
$z_j=\zeta_j+\alpha$ ($j=0,1,2$), 
$\zeta_0=-iX_{\epsilon}$, 
$\zeta_1=iX_{\epsilon+\omega}$, 
$\zeta_2=-s iX_{\epsilon-\omega}$, 
$F^s_{A}=[2z_0(z_0+z_1+z_2)^2+z_1z_2(z_1+z_2)]
[\epsilon^4+6\epsilon^2\Delta^2+\Delta^4
-\omega^2(\epsilon^2+\Delta^2)]/\zeta_1\zeta_2\zeta_0^2$, 
$F^s_{AB}=[2z_0z_1z_2+z_1z_2(z_1+z_2)](\epsilon^2-\omega^2-\Delta^2)
/\zeta_1\zeta_2$, 
$F^s_{B}=-2z_0(z_0^2-z_1z_2)$, and 
$F^s_{C}=2[z_0z_1(2z_0+z_1+z_2)][\epsilon(\epsilon+\omega)+\Delta^2]
/\zeta_1\zeta_0$. 
These terms correspond to various matrix elements, and 
the term with $F^s_{AB}$ causes a finite 
absorption at the band edge $\omega=\Delta$. 
(The other terms have vanishing absorption edges.) 
This behavior is different from that of 
the linear response in which 
the absorption vanishes at the band edge ($\omega=2\Delta$) 
owing to the symmetry $v_k=-v_{-k}$. 
The nonlinear process changes this behavior 
because of the existence of intermediate states. 
The dependence of the absorption on $\alpha$ 
is significant compared with that of the linear response, and 
$\text{Re}\sigma^{(3nv)}\propto 1/\alpha^3$ 
for $\alpha\gg\Delta$. 
Therefore, $\text{Re}\sigma^{(3nv)}$ 
is negligible in the dirty limit as compared with 
terms including vertex corrections calculated below.

\section{Vertex Correction}

\begin{figure}
\includegraphics[width=11.5cm]{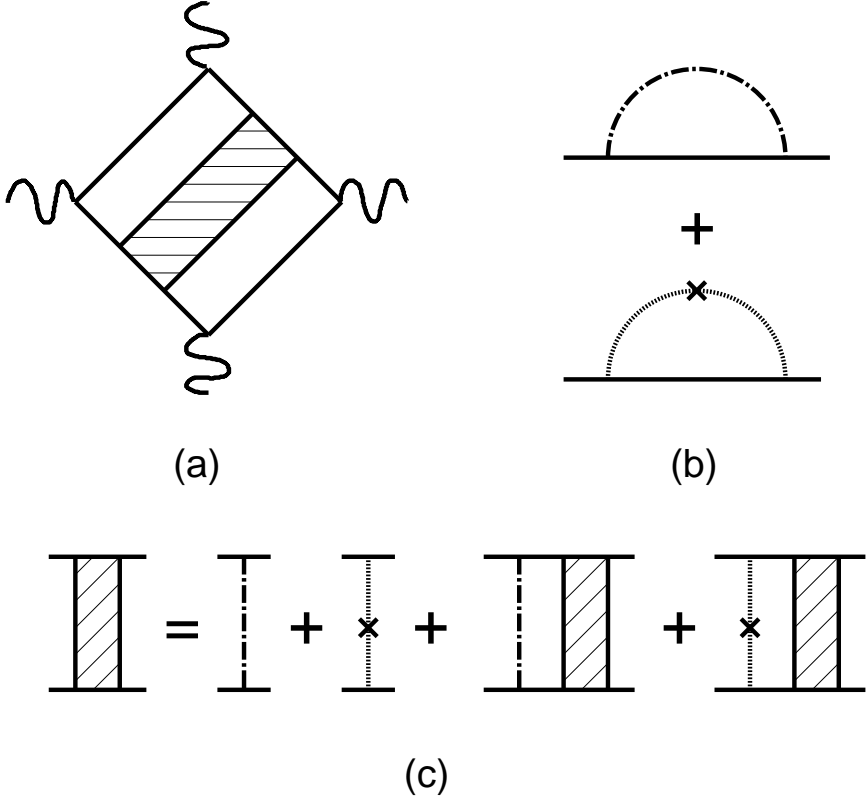}
\caption{\label{fig:1} (a) Diagram of a third-order nonlinear 
response function with vertex corrections. 
The wavy lines, solid lines, and 
shaded square represent 
the external field, electrons, and 
vertex correction, respectively. 
(b) Diagram of the self-energy. 
The dashed line represents the electron$-$phonon interaction. 
The dotted line with a cross 
indicates the impurity scattering. 
(c) Diagrammatic representation of 
the integral equation for the vertex correction.}
\end{figure}

\begin{figure}
\includegraphics[width=11.5cm]{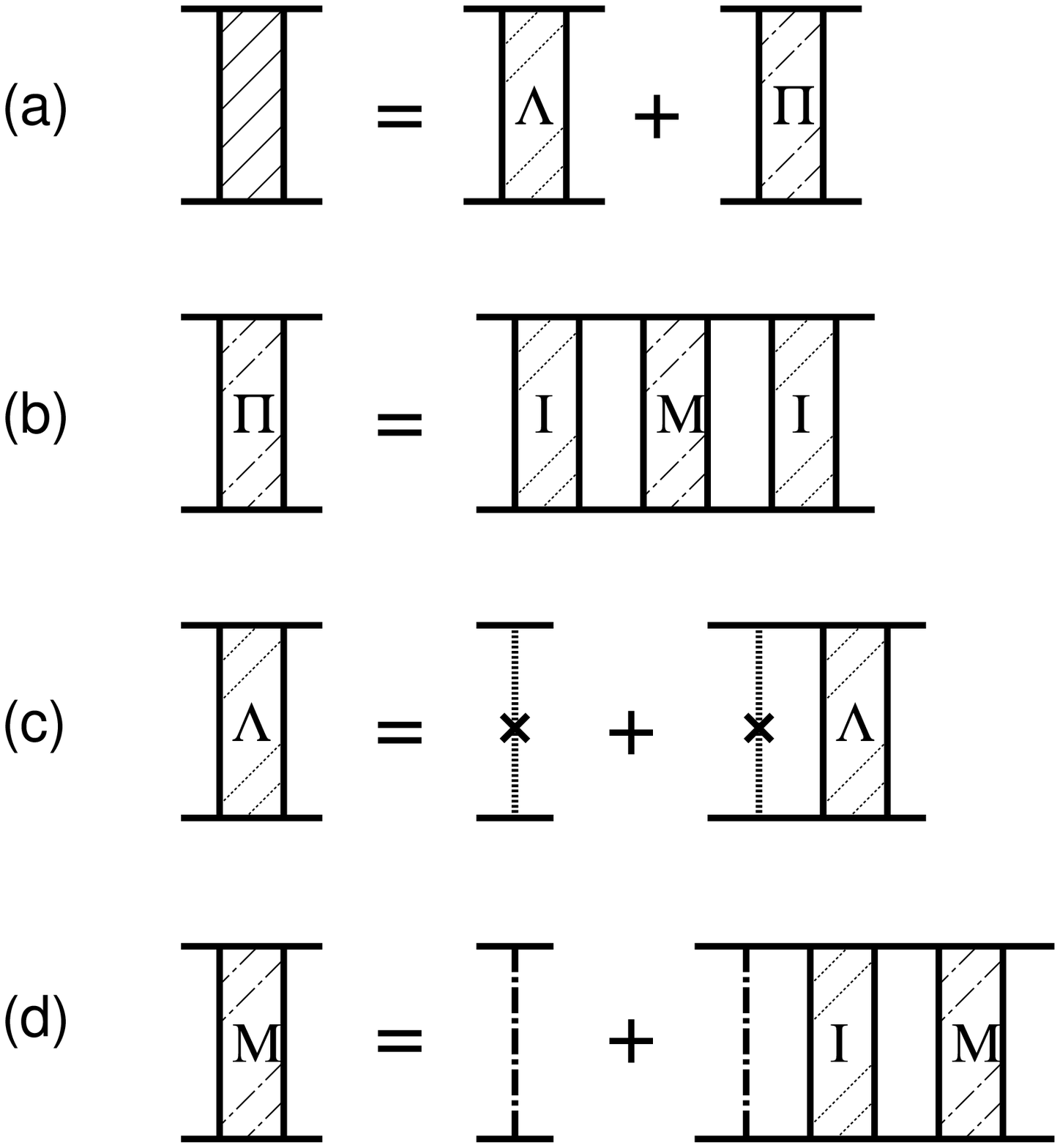}
\caption{\label{fig:2} 
(a) Division of 
the four-point vertex into two terms ($\Lambda$ and $\Pi$). 
(b) Rewriting the term that contains the electron$-$phonon 
interaction as a four-point vertex. 
(c) Integral equation for the vertex correction 
by the impurity scattering. 
(d) Integral equation for the vertex correction 
by the electron$-$phonon interaction. 
The definition of $I$ is given in Sect. 4.1 
as $\check{I}$ [Eq. (\ref{eq:impsct})], and 
$\Lambda$ is equal to $\check{I}-\check{1}$. 
The expression for $\check{M}$ is also given in Sect. 4.1 
[Eq. (\ref{eq:melph})].} 
\end{figure}

The current density with the vertex correction 
is also derived from Eq. (\ref{eq:crnt}). 
The diagram of the response function is shown in 
Fig.~\ref{fig:1}(a). 
The vertex correction 
is introduced, consistent 
with the conserving approximation.~\cite{baym} 
The irreducible four-point vertex is 
obtained from the self-energy. 
The latter is already included 
in the above calculations and its diagram is shown in 
Fig. 1(b). 
From this interaction term, 
the vertex correction is obtained by solving an integral equation 
with its diagrammatic representation given in 
Fig. 1(c). 
We divide this into two terms as in 
Fig.~\ref{fig:2}(a); 
one contains the electron$-$phonon interaction [Figs. 2(b) and 2(d)]
and the other does not contain it as a four-point vertex [Fig. 2(c)]. 
The former is called a vertex correction 
by the amplitude mode 
because it leads to the variation of the superconducting gap 
as shown below, 
and the response function is written as $K^{(3am)}_{\omega}$, 
which contains $\Pi$ of Fig. 2(b) as the vertex correction. 
The latter is the vertex correction 
by the impurity scattering 
because it contains only the impurity scattering 
as a four-point vertex, and 
the response function is written as $K^{(3im)}_{\omega}$, 
which contains $\Lambda$ of Fig. 2(c) as the vertex correction.  
Then, the response function with the vertex correction 
is written as 
$K^{(3vc)}_{\omega}=K^{(3am)}_{\omega}+K^{(3im)}_{\omega}$. 

\subsection{Vertex correction by the amplitude mode}

First, we show the calculation of $K^{(3am)}_{\omega}$. 
The weak coupling approximation is introduced 
for the electron$-$phonon interaction 
as in the case of the self-energy. 
(This approximation corresponds to neglecting 
the inelastic scattering, which is considered to 
be small at low temperatures.) 
Then, the result is written as 
\begin{equation}
 K^{(3am)}_{\omega}=
\frac{\text{e}^4|E|^2}{2\omega^2}\frac{g_{ph}^2}{\omega_E}
{}^t\bm{\chi}^{(S)}_{\omega}
\check{\cal T}
\check{M}_{\omega}\bm{\chi}^{(K)}_{\omega}\theta^<_{\omega}. 
\label{eq:k3am1}
\end{equation}
Here, 
$\check{\cal T}=\bigl(\begin{smallmatrix}
\hat{1} & \hat{0} \\ \hat{0} & -\hat{\tau}_1 \end{smallmatrix}\bigr)$, 
$\bm{\chi}_{\omega}^{(K)}=
\sum_{s=\pm}s\left(
\bm{h}^{++s}_{-\omega}+\bm{h}^{+s-}_{0}+\bm{h}^{s--}_{\omega}\right)$, 
and 
$\bm{\chi}_{\omega}^{(S)}
=\sum_{s,u=\pm}s\left(
\bm{h}^{+us}_{-\omega}+\bm{h}^{usu}_{0}+\bm{h}^{su-}_{\omega}
\right)$. 
The vector ${}^t\bm{h}=(h_{11},h_{22},
h_{12},h_{21})$ is made from the corresponding 
components of a $2\times 2$ matrix $\hat{h}$ 
(${}^t$ indicates a transposition). 
\[
\begin{split}
 (\hat{h}^{abc}_x)_{i,j}=
\frac{1}{N^3}\sum_kv_k^2 &
\int\frac{\text{d}\epsilon}{2\pi i}
\text{tanh}\left(\frac{\epsilon+x}{2T}\right) 
\sum_{n,l}
\left(\check{I}^{a,c}_{\epsilon+\omega,\epsilon-\omega}\right)_{n,l}^{i,j} \\
&\times
[\hat{\tau}_3\hat{G}^a_k(\epsilon+\omega)
\hat{G}^b_k(\epsilon)\hat{G}^c_k(\epsilon-\omega)\hat{\tau}_3]_{n,l} 
\end{split}
\]
($a,b,c=+$ or $-$ and $i,j,n,l=1,2$ designate the 
elements of a matrix) 
with the vertex correction by the impurity scattering; 
\begin{equation}
\check{I}^{a,c}_{\epsilon+\omega,\epsilon-\omega}
=(\check{1}-n_iu^2
\check{m}_{\epsilon+\omega,\epsilon-\omega}^{a,c})^{-1}. 
\label{eq:impsct} 
\end{equation}
Here, $\check{1}$ is a $4\times 4$ unit matrix 
and 
\[
(\check{m}_{\epsilon+\omega,\epsilon-\omega}^{a,c})_{n,l}^{i,j}=
\frac{1}{N^3}\sum_k
[\hat{\tau}_3\hat{G}^a_k(\epsilon+\omega)]_{i,n}
[\hat{G}^c_k(\epsilon-\omega)\hat{\tau}_3]_{l,j}. 
\]
The vertex correction by the electron$-$phonon interaction is 
\begin{equation}
\check{M}_{\omega}
=\left(
\check{1}
-\check{R}^{+,+}_{-\omega}+\check{R}^{+,-}_{-\omega}
-\check{R}^{+,-}_{\omega}+\check{R}^{-,-}_{\omega}
\right)^{-1}
\label{eq:melph}
\end{equation}
with 
\[
\check{R}^{a,b}_{x}=
\frac{g_{ph}^2}{\omega_E}
\int\frac{\text{d}\epsilon}{2\pi i}
\text{tanh}\left(\frac{\epsilon+x}{2T}\right)
\check{I}^{a,b}_{\epsilon+\omega,\epsilon-\omega}
\check{m}_{\epsilon+\omega,\epsilon-\omega}^{a,b}. 
\]
Equation (\ref{eq:k3am1}) is rewritten as 
$K^{(3am)}_{\omega}=
(\text{e}^2/2){}^t\bm{\chi}^{(S)}_{\omega}
\check{\cal T}
\bm{\delta\Delta}_{\omega}\theta^<_{\omega}$. 
Here, 
$\bm{\delta\Delta}_{\omega}=
(g_{ph}^2/\omega_E)(\text{e}|E|/\omega)^2
\check{M}_{\omega}\bm{\chi}^{(K)}_{\omega}$ 
is interpreted as a nonequilibrium part of the self-energy 
because it includes the effect of the external field. 
$\hat{\chi}^{(K,S)}_{\omega}$ and $\hat{\delta\Delta_{\omega}}$ 
are proportional to $\hat{\tau}_1$ because of 
the particle$-$hole symmetry, and 
$\hat{\tau}_1$ represents an off-diagonal order 
as is known from expressions of the Green's function of electrons 
and the gap equation. 
Therefore, 
$\bm{\delta\Delta}_{\omega}$ represents the 
variation of the superconducting gap under the external 
field, and 
the equation for $\bm{\delta\Delta}_{\omega}$ is 
rewritten as 
$\bm{\delta\Delta}_{\omega}=
(\text{e}|E|/\omega)^2
\bm{\chi}^{(K)}_{\omega}/\Gamma^{(am)}_q(2\omega)$. 
Here, $1/\Gamma^{(am)}_q(2\omega)=
[(\check{M}_{\omega})_{1,2}^{1,2}
+(\check{M}_{\omega})_{2,1}^{1,2}]
g_{ph}^2/\omega_E$, and 
$\Gamma^{(am)}_q(2\omega)$ represents an amplitude mode.~\cite{littlewood} 
This is calculated as 
\begin{equation}
\begin{split}
 \Gamma^{(am)}_q(2\omega)=&
\frac{mk_F}{2\pi^2}\int\text{d}\epsilon
\text{tanh}\left(\frac{\epsilon}{2T}\right) 
\Biggl[
\frac{1}{X^{(r)}_{\epsilon}}  \\
&-
\sum_{s=\pm 1}
\frac{f_s(q)}{1-2 i\alpha f_s(q)}
\left(
\frac{\epsilon(\epsilon+2\omega)+\Delta^2}
{X_{\epsilon+2\omega}X^{(r)}_{\epsilon}}
+s\right)
\Biggr]. 
\label{eq:ampfl}
\end{split}
\end{equation}
Here, 
$f_s(q)=2^{-1}\int_{-1}^1\text{d}(\text{cos}\theta)
/(X_{\epsilon+2\omega}+sX^{(r)}_{\epsilon}+2 i\alpha
-v_Fq\text{cos}\theta)$ 
and $X^{(r)}_{\epsilon}=\text{Re}X_{\epsilon}$. 

In TPA, there exists another type of coupling 
vertex with external fields: 
$\hat{\tau}_3\partial v_k/\partial k$ 
(the diamagnetic term). 
In this case, 
the quantity corresponding to $\hat{\chi}^{(K)}_{\omega}$ 
is proportional to linear superpositions of 
$\hat{\tau}_3$ and $\hat{\tau}_2$. 
This represents density ($\hat{\tau}_3$) 
and phase ($\hat{\tau}_2$) fluctuations, 
and these result in a plasma oscillation. 
This term is negligible because of its high-energy 
excitation energy and the smallness of this type of vertex.~\cite{note2} 
In the case of the linear response, 
the quantity corresponding to $\hat{\chi}^{(K)}_{\omega}$ 
includes the single vertex $v_k$ in contrast to 
$v_k^2$ in TPA. Therefore, the amplitude mode 
does not contribute to the linear absorption 
because of the inversion symmetry $v_{-k}=-v_k$. 
This holds true even if the nonlocality 
($q\ne 0$) is introduced.~\cite{note3}

The resulting absorption spectrum is given by 
\begin{equation}
 \text{Re}\sigma^{(3am)}_{\omega}=
\sigma_{\Delta}
\left(\frac{\text{e}\xi_0|E|}{\omega}\right)^2
\frac{2\Delta}{\omega}
\text{Im}\left[
\frac{\chi'^2_{\omega}}
{\Gamma'_{q=0}(2\omega)}
\right]\theta^<_{\omega}. 
\label{eq:ampab}
\end{equation}
Here, 
$\Gamma'_{q}(2\omega)=(mk_F)^{-1}
\Gamma^{(am)}_{q}(2\omega)$ and 
$\chi'_{\omega}=m\Delta k_F^{-3}
(\hat{\chi}^{(K)}_{\omega})_{1,2}$ are 
dimensionless quantities. 
($\hat{\chi}^{(K)}_{\omega}=\hat{\chi}^{(S)}_{\omega}/2$ 
at $T=0$.) 
From Eq. (\ref{eq:ampfl}), 
$\text{Re}\Gamma'_{q=0}(2\omega)=
(1/\pi^2)\sqrt{1-\Delta^2/\omega^2}
\text{arcsinh}\sqrt{(\omega/\Delta)^2-1}$ 
and 
$\text{Im}\Gamma'_{q=0}(2\omega)=
-(1/2\pi)\sqrt{1-\Delta^2/\omega^2}$ 
for $\omega\ge\Delta$. 
This expression of $\Gamma'_{q=0}(2\omega)$ 
shows that, in the local limit, 
the amplitude mode is independent of 
the impurity scattering 
when its vertex correction is included, 
consistent with a conservation law. 
This property is similar to the fact that 
nonmagnetic impurities are ineffective 
on the gap in conventional $s$-wave 
superconductors.~\cite{anderson2} 
This result is consistent with 
an observation of 
the long-time ($\gg\tau$) oscillation 
of the amplitude mode,~\cite{matsunaga} 
and gives an explanation for this behavior. 

Although the amplitude mode is independent of $\alpha$, 
the absorption by this mode depends on the impurity scattering. 
In Eq. (\ref{eq:ampab}), 
$\chi'$ is roughly proportional to $1/\alpha$ 
for $\alpha\gg\Delta$; 
\[
\text{Re}\chi'_{\omega}\simeq 
\frac{-\Delta^2}{6\pi^2\alpha}
\int_{\omega-\Delta}^{\infty}\text{d}\epsilon
\text{Im}
\left(
\frac{3\epsilon^2-\omega^2+\Delta^2+X_{\epsilon+\omega}X_{\epsilon-\omega}}
{X_{\epsilon}X_{\epsilon+\omega}X_{\epsilon-\omega}
(X_{\epsilon+\omega}+X_{\epsilon-\omega})}\right)
\]
and 
$\text{Im}\chi'_{\omega}\simeq
-(\Delta/12\pi\alpha)
\sqrt{1-\Delta^2/\omega^2}$. 
Then, 
$\text{Re}\sigma^{(3am)}\propto 1/\alpha^2$. 
The effect of the amplitude mode on 
the absorption spectrum 
is clearly observed at the band edge, 
which is 
$\text{Re}\sigma^{(3am)}\propto 1/\sqrt{\omega-\Delta}$ 
around $\omega\simeq \Delta$ 
because 
$\text{Re}\Gamma'_{q=0}(2\omega)\propto \omega-\Delta$, 
$\text{Im}\Gamma'_{q=0}(2\omega)\propto \sqrt{\omega-\Delta}$, 
and $\chi'_{\omega}\simeq \text{Re}\chi'_{\Delta}\ne 0$ 
for $\omega\to \Delta$. 
The amplitude mode appears not around $\omega=2\Delta$ 
but around $\omega=\Delta$. This is because 
the TPA results from two photons with the frequency 
of a photon being $\omega$, and 
$\omega=\Delta$ means $2\omega=2\Delta$. 
Here, ``two photons'' does not indicate the 
quantized electromagnetic field, but 
the second order of the external field 
(the excitation by $A_{\omega}^2$). 

\subsection{Vertex correction by the impurity scattering}

$K^{(3im)}_{\omega}$ is derived in a similar way. 
In the case of $\omega<2\Delta$ and $T\to 0$, 
the finite contribution to $\text{Im}K^{(3)}$ 
comes from the term that includes a factor such as 
$\text{tanh}(\epsilon_-/2T)-
\text{tanh}(\epsilon_+/2T)$ 
($\epsilon_{\pm}=\epsilon\pm\omega$). 
Then, the calculated result for the absorption spectrum 
is written as 
\begin{equation}
 \text{Re}\sigma^{(3im)}_{\omega}=
\sigma_{\Delta}
\left(\frac{\text{e}\xi_0|E|}{\omega}\right)^2
\frac{\alpha\Delta^3(2F'_{\omega}+F''_{\omega})}
{9\pi^2\omega}\theta^<_{\omega}. 
\label{eq:impab}
\end{equation}
Here, 
$F'_{\omega}=
\int_{\Delta-\omega}^{-\Delta+\omega}
\text{d}\epsilon
{}^t\bm{\kappa}^{+i+}_{\epsilon,\epsilon+\omega,\epsilon}
\check{\cal T}
\check{I}^{+,+}_{\epsilon,\epsilon}
\bm{\kappa}^{+i+}_{\epsilon,\epsilon-\omega,\epsilon}$ 
and 
$F''_{\omega}=-\int_{\Delta-\omega}^{-\Delta+\omega}
\text{d}\epsilon
\sum_{s=\pm}s
\text{Re}(
{}^t\bm{\kappa}^{sr+}_{\epsilon_-,\epsilon,\epsilon_+}
\check{\cal T}
\check{I}^{+,s}_{\epsilon_+,\epsilon_-}
\bm{\kappa}^{+rs}_{\epsilon_+,\epsilon,\epsilon_-})$ 
with 
$\hat{\kappa}^{s_1 c s_2}_{\epsilon_1,\epsilon_2,\epsilon_3}=
\int\text{d}\xi_k
\hat{\tau}_3
\hat{G}^{s_1}_{k}(\epsilon_1)
{\cal C}
[\hat{G}^+_k(\epsilon_2)]
\hat{G}^{s_2}_{k}(\epsilon_3)\hat{\tau}_3/\pi$ 
(${\cal C}=\text{Re}$, $\text{Im}$ for $c=r$, $i$, respectively). 
The analytic calculation shows that 
$\text{Re}\sigma_{\omega}^{(3im)}\propto 1/\alpha^2$ 
for $\alpha\gg\Delta$. 
The absorption edge at $\omega=\Delta$ is 
finite as in the case of 
$\text{Re}\sigma_{\omega}^{(3nv)}$.

\section{Numerical Calculations}

\begin{figure}[here]
\includegraphics[width=11.5cm]{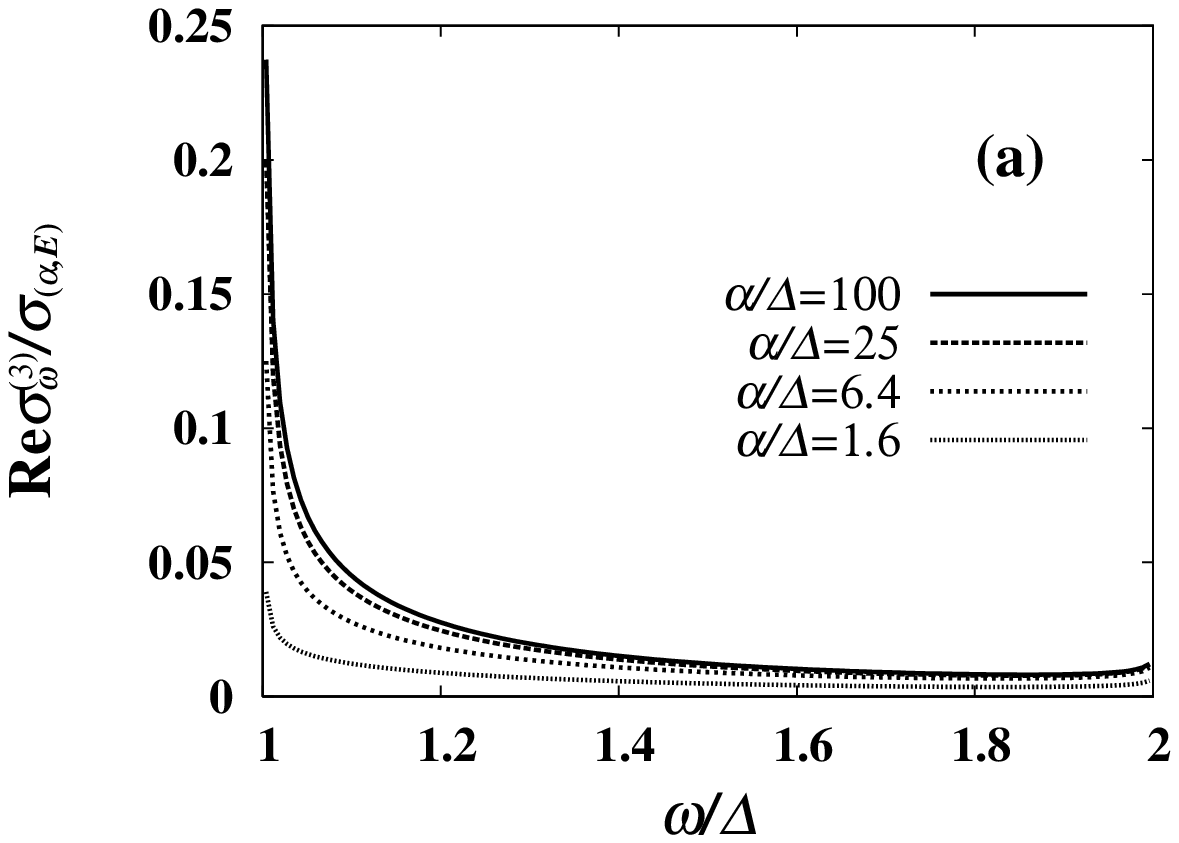}
\includegraphics[width=11.5cm]{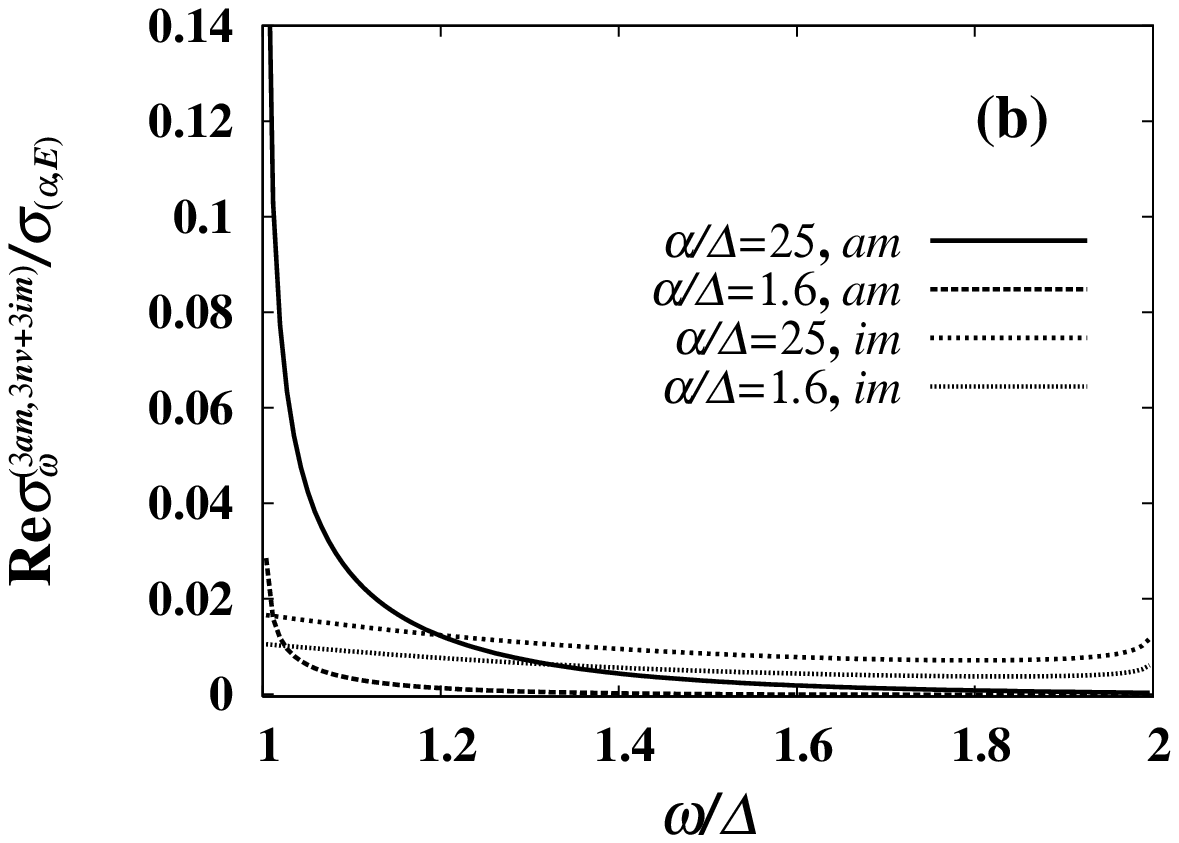}
\caption{(a) Dependence of 
$\text{Re}\sigma_{\omega}^{(3)}/\sigma_{(\alpha,E)}$ 
on $\omega$ for several values of $\alpha$. 
(b) Dependences of 
$\text{Re}\sigma_{\omega}^{(3am)}/\sigma_{(\alpha,E)}$ (noted as $am$) 
and 
$(\text{Re}\sigma_{\omega}^{(3nv)}+\text{Re}\sigma_{\omega}^{(3im)})
/\sigma_{(\alpha,E)}$ 
(noted as $im$) 
on $\omega$ for $\alpha/\Delta=1.6$ and $25$.} 
\label{fig:3}
\end{figure}

Numerical calculations 
are carried out 
for Eqs. (\ref{eq:linab}), (\ref{eq:novcab}), 
(\ref{eq:ampab}), and (\ref{eq:impab}) 
for general values of $\alpha$ 
without taking $\alpha\gg\Delta$. 
The results of numerical calculations 
for $\text{Re}\sigma_{\omega}^{(3)}/\sigma_{(\alpha,E)}$ are shown in 
Fig.~\ref{fig:3}(a). 
Here, 
$\text{Re}\sigma_{\omega}^{(3)}=
\text{Re}\sigma_{\omega}^{(3nv)}+
\text{Re}\sigma_{\omega}^{(3am)}+
\text{Re}\sigma_{\omega}^{(3im)}$ and 
$\sigma_{(\alpha,E)}
=\sigma_{\Delta}(\text{e}\xi_0|E|/\alpha)^2$. 
As $\text{Re}\sigma_{\omega}^{(3)}\propto 1/\alpha^2$, 
$\text{Re}\sigma_{\omega}^{(3)}/\sigma_{(\alpha,E)}$ becomes 
independent of $\alpha$ for $\alpha\gg\Delta$. 
Although this behavior is indicated by 
an analytical calculation, 
numerical calculations show that 
$\text{Re}\sigma_{\omega}^{(3am)}$ 
becomes predominant over $\text{Re}\sigma_{\omega}^{(3im)}$ 
for large values of $\alpha$, as shown in 
Fig.~\ref{fig:3}(b). 
This originates from 
the $\omega$-dependence of $\chi'_{\omega}$, 
which varies with $\alpha$. 

\begin{figure}[here]
\includegraphics[width=11.5cm]{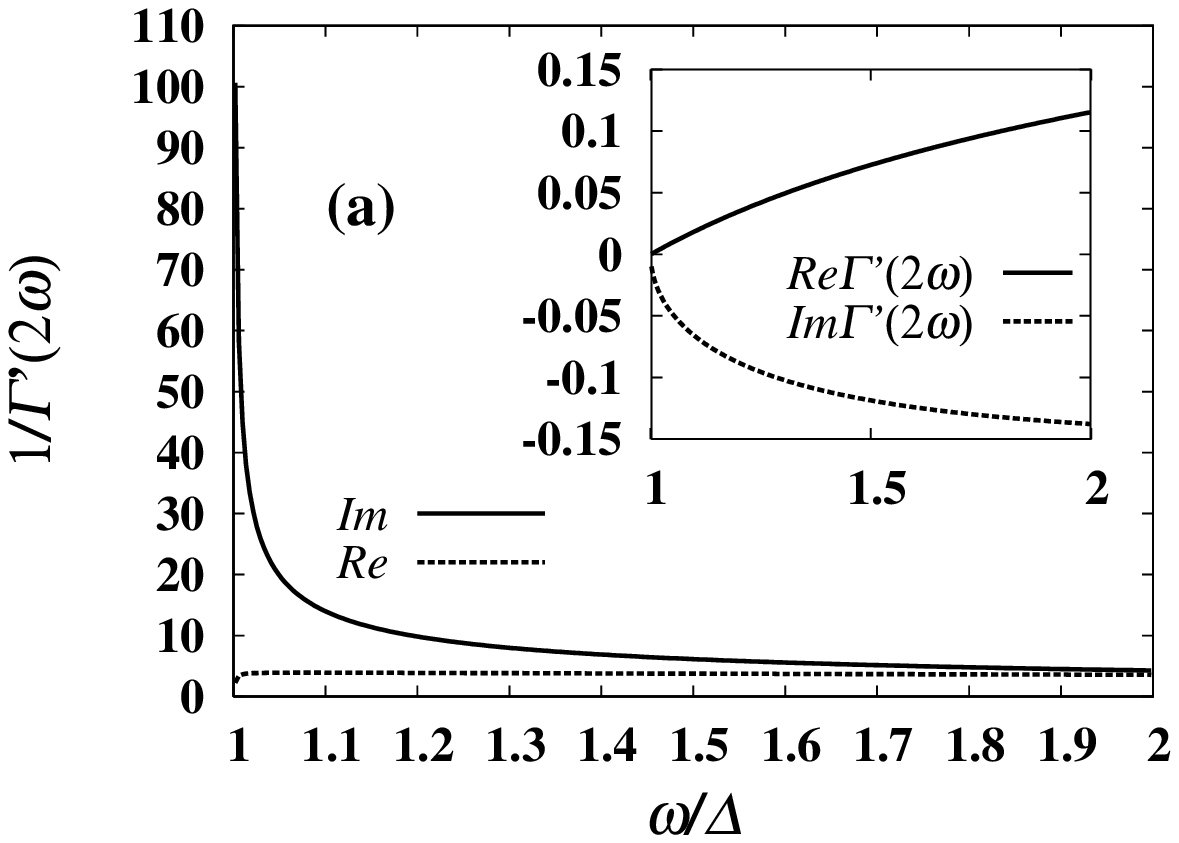}
\includegraphics[width=11.5cm]{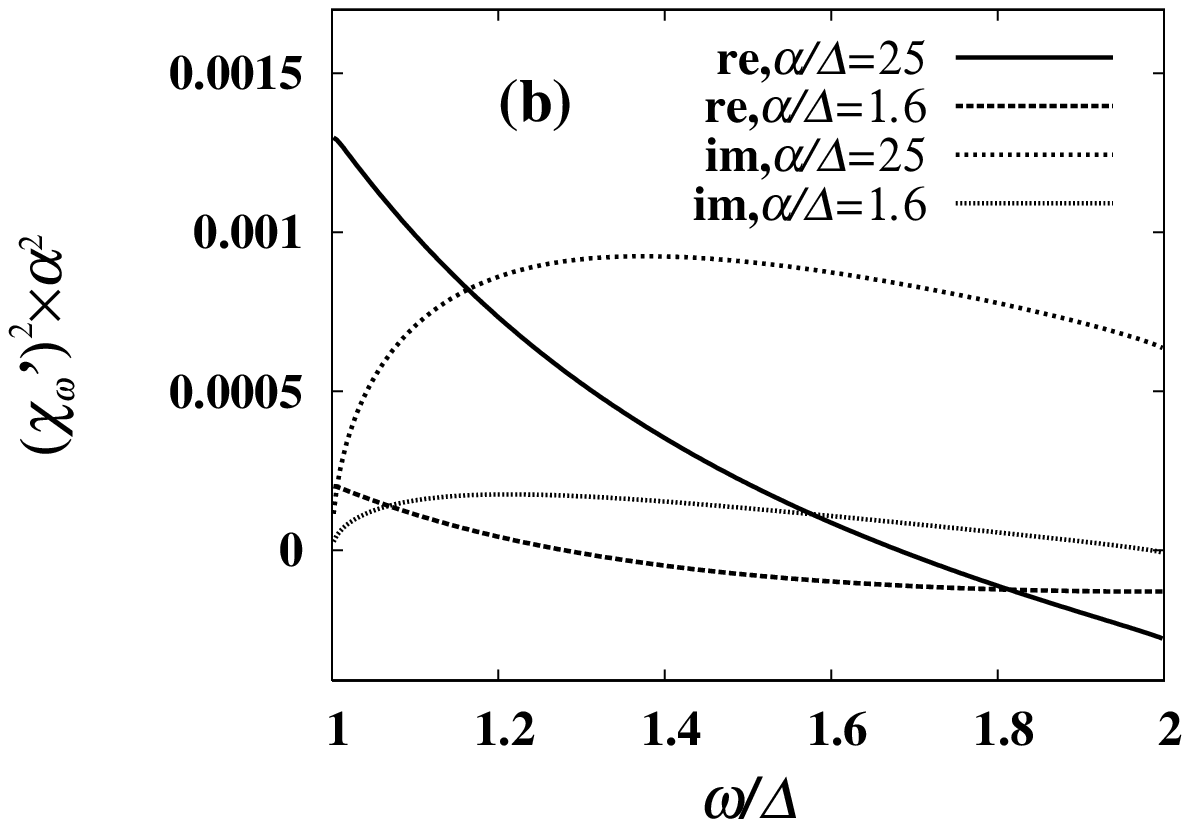}
\caption{(a) Dependences of 
the real (noted as $Re$) and imaginary ($Im$) parts 
of $1/\Gamma'_{q=0}(2\omega)$ 
on $\omega$. 
The inset shows $\omega$-dependences of 
$\text{Re}[\Gamma'_{q=0}(2\omega)]$ 
and $\text{Im}[\Gamma'_{q=0}(2\omega)]$. 
(b) Dependences of 
the real (noted as re) and imaginary (im) parts 
of $\alpha^2\chi'^2_{\omega}$ 
on $\omega$ for $\alpha/\Delta=1.6$ and $25$.} 
\label{fig:4}
\end{figure}

The dependences of $1/\Gamma'_{q=0}(2\omega)$ 
and $\chi'^2_{\omega}$ [in Eq. (\ref{eq:ampab})] 
on $\omega$ are 
shown in Figs.~\ref{fig:4}(a) and 4(b) 
to clarify the contribution of the amplitude mode 
to the TPA spectrum. 
These dependences of $\Gamma'$ and $\chi'$ 
on $\omega$ are the same as expected in Sect. 4.1. 
$\text{Im}[1/\Gamma'_{q=0}(2\omega)]$ indicates 
the spectral weight of the amplitude mode, 
and 
$\text{Re}\sigma_{\omega}^{(3am)}\propto 
\text{Im}(\chi'_{\omega})^2
\text{Re}[1/\Gamma'_{q=0}(2\omega)]
+\text{Re}(\chi'_{\omega})^2
\text{Im}[1/\Gamma'_{q=0}(2\omega)]$ 
from Eq. (\ref{eq:ampab}).  
This shows that 
the predominant contribution 
to the TPA spectrum around $\omega\simeq\Delta$ 
results from the amplitude mode 
because $\text{Re}(\chi'_{\omega})^2$ 
is larger than $\text{Im}(\chi'_{\omega})^2$ 
for this region. 
For $\omega<\Delta$, 
$\text{Im}\Gamma'_{q=0}(2\omega)=0$ 
and 
$\text{Im}\chi'_{\omega}=0$ 
because of the absence of thermal excitation in the 
low-temperature limit, and then 
the TPA vanishes in this region. 
In the local limit ($q=0$), 
the dependence of $\Gamma'_q(2\omega)$ on $q$ 
(the dispersion relation of the amplitude mode) 
is irrelevant to the TPA spectrum. 

\begin{figure}[here]
\includegraphics[width=11.5cm]{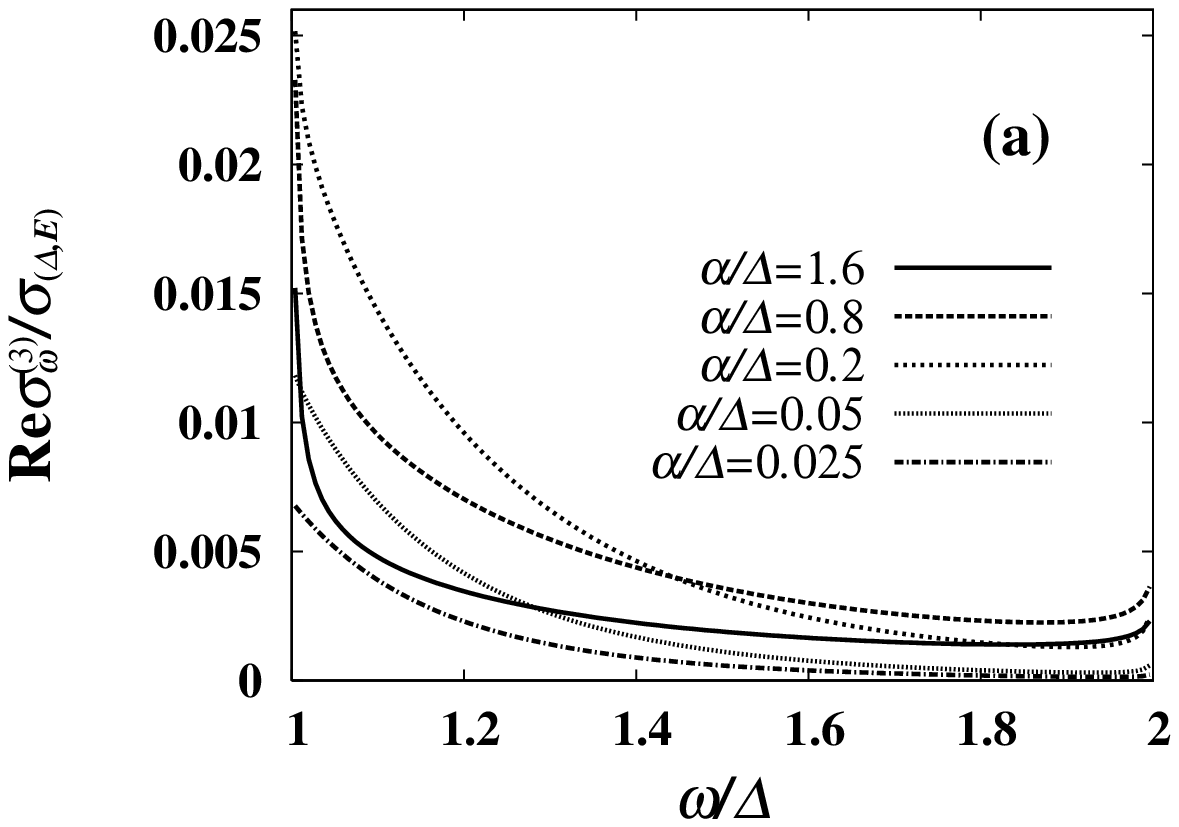}
\includegraphics[width=11.5cm]{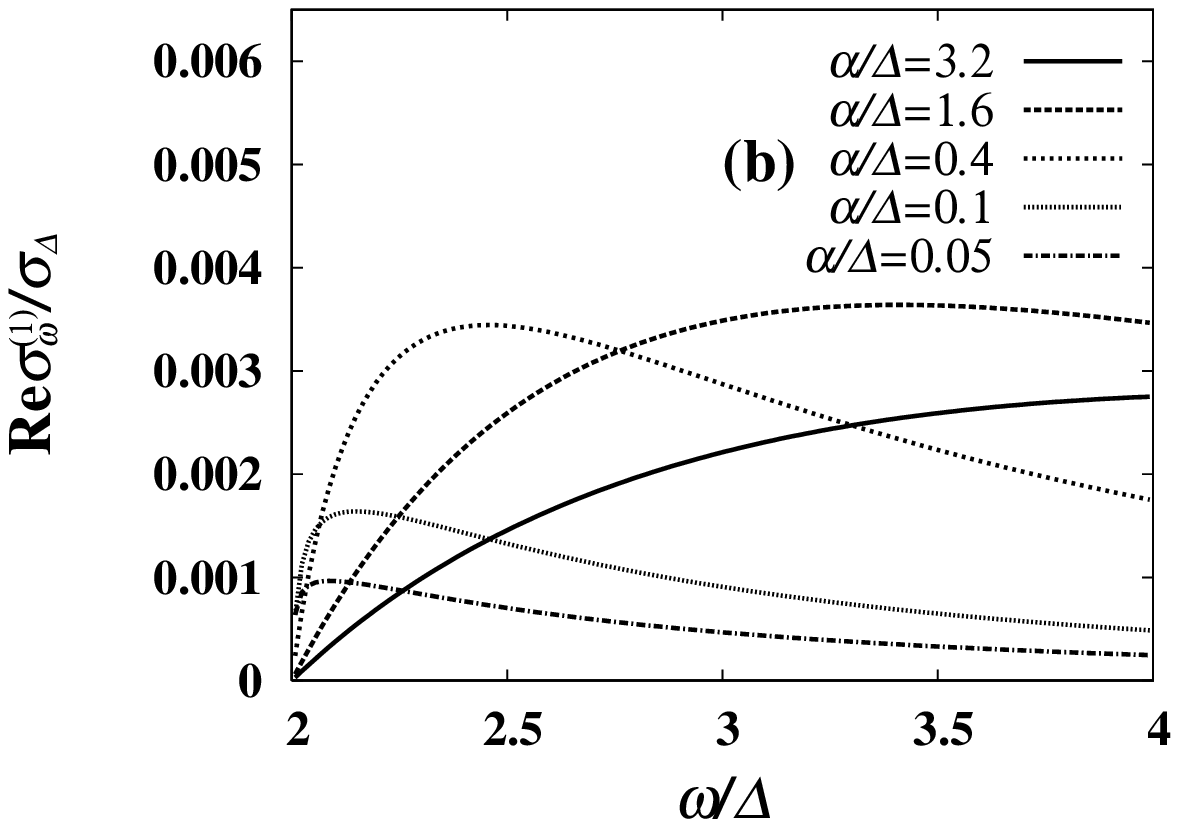}
\caption{Dependences of 
(a) $\text{Re}\sigma_{\omega}^{(3)}/\sigma_{(\Delta,E)}$ and 
(b) $\text{Re}\sigma_{\omega}^{(1)}/\sigma_{\Delta}$ 
on $\omega$ for several values of $\alpha$.
} 
\label{fig:5}
\end{figure}

The variation of $\text{Re}\sigma_{\omega}^{(3)}$ 
from the clean ($\alpha/\Delta\ll 1$) to dirty ($\alpha/\Delta>1$) cases 
is shown in Fig.~\ref{fig:5}(a). 
[$\sigma_{(\Delta,E)}=\sigma_{\Delta}(\text{e}\xi_0|E|/\Delta)^2$ 
in the denominator is independent of $\alpha$.] 
There is no absorption in the clean limit ($\alpha/\Delta\to 0$) 
as noted above,~\cite{note1} 
and the TPA increases with increasing $\alpha$ in the 
case of $\alpha/\Delta\ll 1$. 
In the dirty case, the TPA decreases with increasing 
$\alpha$ as noted in previous sections. 
The reason for this is that the absorption 
occurs in the wider $\omega$ range 
as $\alpha$ increases, which is also seen 
in the linear absorption in Fig. 5(b). 
This shift of the spectral weight 
contributes to a relative predominance 
of the amplitude mode in the TPA 
around $\omega\simeq \Delta$ 
for the dirty limit. 

We quantitatively estimate 
$\text{Re}\sigma_{\omega}^{(3)}/\sigma_0$, which 
is equal to 
$6\pi^2[(\text{e}\xi_0|E|)^2/\alpha\Delta]
(\text{Re}\sigma_{\omega}^{(3)}/\sigma_{(\alpha,E)})$ 
and proportional to $|E|^2/\alpha$ for 
$\alpha\gg\Delta$. 
If we put $|E|=5$ kV/cm, $\xi_0/\pi=5$ nm, $\Delta=2.5$ meV, 
and $\alpha/\Delta=100$, then 
$(\text{Re}\sigma_{\omega}^{(3)}/\sigma_0)\simeq 
5.8\times 
(\text{Re}\sigma_{\omega}^{(3)}/\sigma_{(\alpha,E)})$. 
This indicates that the TPA by the amplitude mode 
can be observed in roughly the same magnitude as the 
linear response. 
Because this calculation is based on a perturbative expansion 
by the external field, 
the magnitude of the electric field 
should be adjusted so that 
the $\text{Re}\sigma^{(3)}/\text{Re}\sigma^{(1)}$ ratio 
is smaller than 1 for the validity of this theory. 

\section{Summary and Discussion}

We calculated the TPA 
spectrum in dirty superconductors. 
As for the absorption edge and 
the dependence on the impurity scattering, 
there are differences between 
the linear and nonlinear responses. 
In the dirty limit, 
the term with vertex corrections gives a 
predominant contribution to the TPA. 
Although the finite absorption generally exists at 
the band edge in the TPA spectrum, 
the amplitude mode is clearly observed 
because of a singular behavior 
as $1/\sqrt{\omega-\Delta}$. 
This mode is not suppressed in the 
dirty limit because a self-energy effect
by the impurity scattering is canceled out with 
its vertex correction.  
Our result shows that 
the TPA would provide a probe to observe 
the amplitude mode experimentally. 

Although we consider the steady state, 
our method can be extended to investigate 
the transient state by introducing 
the integration over frequencies of the external field. 
In the latter case, the restriction 
on $\omega$ ($\omega<2\Delta$) is removed 
because of the finite radiation time. 
This type of study would give a direct 
comparison of a theory to the 
pump-probe spectroscopy.~\cite{matsunaga} 

Investigating various types of interaction 
is also a possible application of our method. 
The method of calculating the vertex correction 
with the conserving approximation can be extended 
beyond the mean field approximation 
(for example, strong coupling effect). 
This is in contrast to a method using 
the Hubbard-Stratonovich transformation 
for the attractive interaction. 
For example, 
in some unconventional superconductors such as cuprates, 
there is no gap structure in the linear absorption 
spectrum,~\cite{kamaras} 
unlike in the conventional superconductors studied here. 
This type of system should be 
investigated beyond the mean field approximation, 
as this will be an interesting subject 
for studying the amplitude mode by our method.

\section*{Acknowledgement}

The numerical computation in this work was carried out 
at the Yukawa Institute Computer Facility.


\begin{thebibliography}{9}

\bibitem{schon} For example, 
G. Sch\"on, in {\it Nonequilibrium Superconductivity}, 
ed. D. N. Langenberg and A. I. Larkin 
(Elsevier, New York, 1986) Chap. 13. 

\bibitem{anderson1} P. W. Anderson, 
Phys. Rev. {\bf 112}, 1900 (1958). 

\bibitem{leggett} A. J. Leggett, 
Prog. Theor. Phys. {\bf 36}, 901 (1966). 

\bibitem{jujo12} T. Jujo, 
J. Phys. Soc. Jpn {\bf 81}, 044710 (2012). 

\bibitem{clarke} J. Clarke and F. K. Wilhelm, 
Nature {\bf 453}, 1031 (2008). 

\bibitem{schmid} A. Schmid, 
Phys. Kondens. Mater. {\bf 8}, 129 (1968). 

\bibitem{volkov} A. F. Volkov and S. M. Kogan, 
Sov. Phys. JETP {\bf 38}, 1018 (1974). 

\bibitem{sooryakumar} R. Sooryakumar and M. V. Klein, 
Phys. Rev. Lett. {\bf 45}, 660 (1980).

\bibitem{littlewood} P. B. Littlewood and C. M. Varma, 
Phys. Rev. B {\bf 26}, 4883 (1982). 

\bibitem{matsunaga} R. Matsunaga, Y. I. Hamada, K. Makise, Y. Uzawa, 
H. Terai, Z. Wang, and R. Shimano, 
Phys. Rev. Lett. {\bf 111}, 057002 (2013).

\bibitem{note1}
The absorption spectrum of 
the Mattis-Bardeen~\cite{mattis} type is observed in 
the nonlocal limit ($v_Fq\gg\Delta$) or 
dirty materials ($\alpha\gg\Delta$). 
(In the former and latter cases, 
the absorption occurs because of the nonconservation of the 
momentum in the propagating direction of external field 
and the finite damping rate of electrons, respectively.) 
Here, $1/q$ is the characteristic length of 
the external field (the magnetic field penetration depth or 
thin film thickness). 
Introducing $q\ne 0$ in calculations 
to make the absorption finite~\cite{papenkort,krull} 
corresponds to a nonlocal case. 
The absorption disappears in 
the clean and local limits ($\alpha\ll\Delta$ and $v_Fq\ll\Delta$). 

\bibitem{keldysh} L. V. Keldysh, 
Sov. Phys. JETP {\bf 20}, 1018 (1965). 

\bibitem{kamenev} A. Kamenev and A. Levchenko, 
Adv. Phys. {\bf 58}, 197 (2009). 

\bibitem{note2}
As for the validity of omitting 
a diamagnetic term ($\partial v_k/\partial k$), 
for example, 
the ratio of 
$\sum_k\int\text{d}\epsilon
\text{Tr}[G\tau_3G\tau_3](\partial v_k/\partial k)^2$ to 
$\sum_k\int\text{d}\epsilon\text{Tr}[GGGG](v_k)^4$ 
is proportional to 
$(1/m)^2/(v_F^2/\Delta)^2$. 
Then, replacing $GGv_k^2$ by $G\partial v_k/\partial k$ 
introduces a small factor $\Delta/E_F$ ($E_F=k_F^2/2m$), and 
the diamagnetic term is negligible. 
This is in contrast to the case of Mott insulators~\cite{jujo08} 
in which the magnitude of the gap is 
comparable to the band width and the $\partial v_k/\partial k$ term 
is predominant over the $v_k$ term. 

\bibitem{abrikosov} A. A. Abrikosov and L. P. Gor'kov, 
Sov. Phys. JETP {\bf 8}, 1090 (1959). 

\bibitem{tsuneto} For example, 
T. Tsuneto, {\it Superconductivity and Superfluidity} 
(Cambridge University Press, New York, 1998) Chap. 4. 

\bibitem{mattis} D. C. Mattis and J. Bardeen, 
Phys. Rev. {\bf 111}, 412 (1958). 

\bibitem{zimmermann} W. Zimmermann, E. H. Brandt, M. Bauer, 
E. Seider, and L. Genzel, 
Physica C {\bf 183}, 99 (1991). 

\bibitem{baym} G. Baym and L. P. Kadanoff, 
Phys. Rev. {\bf 124}, 287 (1961). 

\bibitem{note3} In the linear response, 
the quantity corresponding to $\hat{\chi}^K_{\omega}$ 
is written as 
$(2\pi iN^3)^{-1}\sum_{\mib k}\int\text{d}\epsilon
\hat{\tau}_3v^x_{\mib k}[
\hat{G}^+_{\mib k+\mib q/2}(\epsilon+\omega)
\hat{G}^K_{\mib k-\mib q/2}(\epsilon)
+\hat{G}^K_{\mib k+\mib q/2}(\epsilon+\omega)
\hat{G}^-_{\mib k-\mib q/2}(\epsilon)]
\hat{\tau}_3$. 
Here, the vertex correction by the impurity scattering 
$\check{I}$ [Eq. (\ref{eq:impsct})] is omitted 
because this does not affect the conclusion below. 
The direction of the external field ($x$ in $v_{\mib k}^x)$ and 
vector representations for $k$ and $q$ 
are explicitly written. 
The vector 
$\mib{q}$ of the external potential 
$A^x_{\mib q}(\omega)$ 
is perpendicular to the $x$-axis 
for the transverse response, and then 
the above quantity vanishes. 
This argument is the same as the discussion 
about the ineffectiveness 
of the collective mode for the transverse response 
in Ref. 26. 
%Ref.~\cite{schrieffer}

\bibitem{anderson2} P. W. Anderson, 
J. Phys. Chem. Sol. {\bf 11}, 26 (1959). 

\bibitem{kamaras} K. Kamar\'as, S. L. Herr, C. D. Porter, 
N. Tache, D. B. Tanner, S. Etemad, T. Venkatesan, 
E. Chase, A. Inam, X. D. Wu, M. S. Hegde, and B. Dutta, 
Phys. Rev. Lett. {\bf 64}, 84 (1990).

\bibitem{papenkort} T. Papenkort, V. M. Axt, and T. Khun, 
Phys. Rev. B {\bf 76}, 224522 (2007). 

\bibitem{krull} H. Krull, D. Manske, G. S. Uhrig, and A. P. Schnyder, 
Phys. Rev. B {\bf 90}, 014515 (2014). 

\bibitem{jujo08} T. Jujo, 
J. Phys. Soc. Jpn {\bf 77}, 064703 (2008). 

\bibitem{schrieffer} J. R. Schrieffer, 
{\it Theory of Superconductivity} 
(Addison-Wesley, Redwood City, Calif., 1983) revised ed., Chap. 8, p. 238. 
 

\end{thebibliography}
\end{document}